\documentclass[prl,twocolumn,floatfix,showpacs,amsmath,amssymb] {revtex4}
\usepackage{bm}
\usepackage{graphicx}

\begin{document}

\title{On the melting of the nanocrystalline vortex matter in
high-temperature superconductors}

\author{Yadin Y. Goldschmidt and Eduardo Cuansing}
\affiliation{Department of Physics and Astronomy, University of Pittsburgh,
Pittsburgh, Pennsylvania 15260}
\date{\today}
\begin{abstract}
Multilevel Monte Carlo simulations of the vortex matter in the
highly-anisotropic high-temperature superconductor
Bi$_2$Sr$_2$CaCu$_2$O$_8$ were performed. We introduced low
concentration of columnar defects satisfying $B_\phi\le B$. Both the
electromagnetic and Josephson interactions among pancake vortices were
included. The nanocrystalline, nanoliquid and 
homogeneous liquid phases were identified in agreement with
experiments. We observed the two-step melting process and also noted
an enhancement of the structure factor just prior to the melting
transition. A proposed theoretical model is in agreement with our findings.
\end{abstract}

 
\maketitle

Recently there were several experiments reporting on some interesting
features of the vortex matter melting transition in
Bi$_2$Sr$_2$CaCu$_2$O$_8$ (BSCCO) in the present of columnar
defects (CDs) \cite{Banerjee,Menghini,Banerjee2}. In particular, 
these experiments consider the case that the 
density of the CDs does not exceed the density of flux-lines
(FLs). Denoting the ``matching field'' by $B_\phi=n_{cd} \phi_0$, where
$n_{cd}$ is 
the number of CDs per unit area and $\phi_0=hc/2e$ is the flux
quantum, we are particularly interested in $B_\phi\le B$. The CDs are
created artificially by irradiating the sample by highly energetic heavy
ions, like 1 GeV Pb ions, which produce tubes of non-conducting damaged
material. The number of CDs can be controlled by 
the amount of irradiation. We assume that CDs and the applied magnetic
field are aligned with the
c-axis of the single crystal sample. The limit $B\gg B_\phi$ corresponds
to the formation of the Bose 
glass phase at low temperature where vortices are situated on the
randomly placed CDs \cite{Blatter,Nelson,Radzihovsky,Kees}. 
On the other hand, for $B<B_\phi$ the picture that
emerged from experiments \cite{Banerjee,Menghini,Banerjee2} and from 
simulations \cite{Tyagi,Dasgupta} is that
of the ``crystallites in the pores'', also referred to as the ``porous''
vortex solid. At low temperatures a skeleton (or matrix)
of vortices localized on CDs is formed, while the excess (interstitial) FLs
form hexagonal crystallites in the lacuna between CDs. Thus this phase
has a short ranged translational order, which extends to a distance of
the order of a typical pore size. 

As the temperature is increased, and
if the magnetic field is large enough, this heterogeneous structure
melts in two stages: first the crystallites in the pores melt into a
nanoliquid while the skeleton remain intact, and subsequently the
skeleton melts and the liquid becomes homogeneous. When the magnetic
field is lowered (but still greater than $B_\phi$) these two
transitions coincide into one. This is usually associated with a kink
in the first order melting line. The first stage melting of the
crystallites is observed in the experiments as a step in the
equilibrium local magnetization \cite{Pastoriza,Zeldov,Schilling} 
and in the simulations as a
sharp increase in the transverse fluctuations of the FLs, among other
signatures. The second transition was not observed in the
experiments as a similar jump in the magnetization or another
equilibrium property. It was observed in transport measurement \cite{Banerjee2}
involving transport currents with alternating polarity. This
establishes the transition as dynamical in nature. It remains to be
established that this is also an equilibrium thermodynamic
transition. In order to support this assertion  it was argued \cite{Banerjee2}
that the second ``delocalization'' transition is associated with the
restoration of a broken longitudinal gauge symmetry. However, at this
time it cannot be ruled out that the second transition is only a crossover
associated with a gradual equilibrium change over a finite range of
temperatures. 

The aim of this work was to try to observe the sequence of the two
transitions and in particular the two different kinds of liquid phases
in numerical simulations. Some of the  advantages of using Monte Carlo 
simulations are that one can control all the interactions precisely, and
one can take microscopic pictures and measure physical quantities that
are difficult to measure in experiments, like the magnitude of the
transverse fluctuations of FLs or the amount of their entanglement. A
main disadvantage is that the system simulated is small and hence
phase transitions are always broadened and are not as sharp as those
observed in real experiments.

The method we are using is a multilevel Monte Carlo simulation as
reported in earlier publications \cite{Tyagi}. In a highly anisotropic
material like BSSCO the basic degrees of freedom are pancake vortices
rather than string-like FLs. A stack of pancakes make a FL and nearest
neighbor pancakes along the z-direction interact via the Josephson
interaction (here we used the approximation
recently derived in Ref. \onlinecite{Goldschmidt}). In addition
there is a long range electromagnetic interaction among all pairs of
pancakes \cite{Clem}, that is repulsive when the two pancakes reside in the
same plane and attractive otherwise. Periodic boundary conditions are
implemented in all directions. Similar to
simulations of systems of bosons \cite{Ceperley}, we implement permutations
when connecting FLs between the top and bottom planes of the
simulation slab. Most simulations involved $6\times 6 \times 36$
pancakes making up 36 FLs in 36 planes (some simulations involved 64
FLs). Extensive simulations were carried out for the case of
$B_\phi=50$ G. Since we keep the number of FLs fixed at $36$ this means
that the number of CDs vary as a function of the magnetic field such
that $N_{cd}=36~B_\phi /B$. The simulations were carried out for
fields $B=50$ G to $200$ G and thus $N_{cd}/N_{FL}$ vary between $100\%$ to
$25\%$. We simulated between $T=60$ K to $79$ K with one degree increments. The
critical temperature for BSCCO was assumed to be $T_c=90$ K. Other
parameters used were $\lambda(0)=1700$ \AA~(penetration depth),
$\xi(0)=30$ \AA~(correlation length), $d=15$ \AA~(CuO$_2$ plane separation),
$\gamma=375$~(anisotropy). We assumed that $\lambda$ and $\xi$ have 
temperature dependence that is proportional to $(1-T/T_c)^{-1/2}$.  
\begin{figure}
\includegraphics[width=0.5\textwidth]{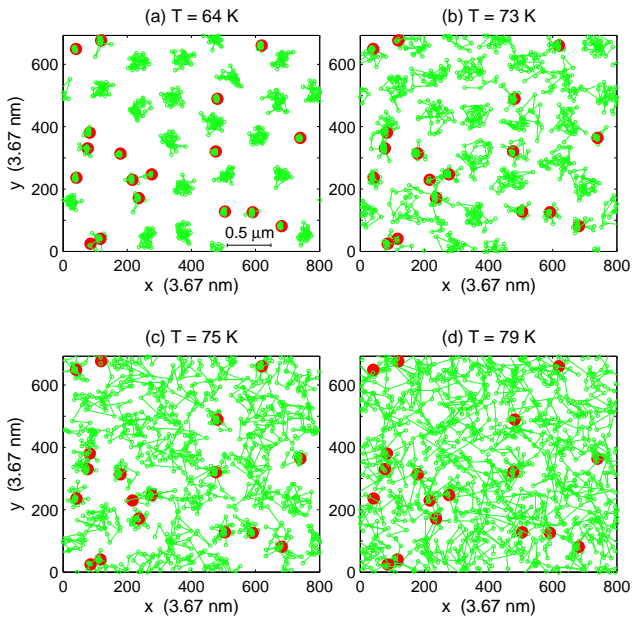}
\caption{(color online) Snapshots of the FLs and CDs (green open circles)
  configurations 
  for $B=100$ G.  The pancakes in 36 planes are projected onto a
  single plane. Pancakes belonging to the same FL are shown to be connected.  
(a) The nanocrystal phase characterized by hexagonal order in the pores.
(b) Just before the melting. A slight enhancement of the structure factor
(see text).
(c) Nanoliquid phase. The pores have melted but not the skeleton.
(d) Homogeneous liquid, full entanglement.}
\label{figure1}
\end{figure}

An important ingredient of the simulation is determining how to model
the interaction between pancakes and CDs in a realistic 
way. There are two major sources of pinning: core and electromagnetic
pinning. Core pinning \cite{Blatter} arises when the vortex core overlaps
with a normal state inclusion similar to the one inside a CD. Since
condensation energy is lost in the vortex core, part or all of this
energy is restored when a vortex resides inside a
CD. Electromagnetic pinning arises \cite{Buzdin,Mkrtchyan} when the
supercurrent pattern around the vortex is disturbed by the
non-conducting defect. These two mechanisms combine
together to yield the expression for the potential energy at a
distance $R$ away from the CD as felt by an individual pancake
\cite{Blatter,Mkrtchyan}: 
\begin{eqnarray}
V(R)\approx \left\{ \begin{array}{c}\epsilon_0 d\
\ln\left[1-\left(\frac{r_r}{R+\xi/\sqrt{2}}\right)^2\right],\ \ R>r_r,
\\
\epsilon_0 d\
\ln\left[1-\left(\frac{r_r}{r_r+\xi/\sqrt{2}}\right)^2\right],\ \ R<r_r,
\end{array}\right.   \label{eq:bind}
\end{eqnarray}
where $\epsilon_0=\phi_0/(4\pi\lambda)^2$ is the energy scale,
and $r_r$ is the radius of the CD. The long range tail contributing $\approx
-\epsilon_0 d r_r^2/R^2$ is due 
mainly to the electromagnetic pinning and the short range flat region of
depth $\approx \epsilon_0 d$ is due to the combined effects of
electromagnetic and core pinning. If $r_r<\sqrt{2}\xi$ a slightly
different expression, 
$V(R)\approx -0.5 \epsilon_0 d r_r^2/(R^2+2\xi^2)$, with a similar long
tail should be used instead \cite{Blatter}. The importance of the long tail was
realized recently \cite{Lopatin} as leading to a dependence $B_{dl}\sim
\exp(-T/T_0)$ instead of $B_{dl}\sim \exp(-T^2/T_0^2)$, leading to the
decoupling of FLs from CDs at a higher temperature than previously thought
\cite{Lopatin}. 
  
After experimenting with CDs of different radii $10$ nm $\le r_r\le
30$ nm we concluded that in 
order to obtain a good agreement with experimental results we needed to
choose $r_r \approx 30$ nm in which case the formula given by
Eq.~(\ref{eq:bind}) is the 
correct one. Different heavy ions with different energies give rise
to tracks of various sizes, normally reported to be in the range of
$4-20$ nm in diameter. It may be that the actual damage caused by a track
exceed its apparent size, thus leading to a higher effective radius.

We now turn to the results of the actual simulations. Most simulations
were done on the Pittsburgh supercomputing cluster and involved
averaging over 10 realizations of the disorder at each
temperature. Typically we ran 4,000,000 Monte Carlo steps for
equilibration and 2,000,000 for measurements. Snapshots are depicted in
Fig.~\ref{figure1} at four representative temperatures. The nanocrystal, (a),
nanoliquid (c) and homogeneous liquid (d) phases are clearly
distinguishable. In addition we display an intermediate picture (b)
taken during the broadened melting transition where the structure
factor slightly 
increases (see later discussion of Fig.~\ref{figure4}) due to a better
rearrangement of the FLs than just before the conclusion of the melting.
In the simulations, when CDs are present, the transition is broadened
probably by the fact that pores of different sizes don't melt
simultaneously especially in a finite size system. 
\begin{figure}
\includegraphics[width=0.5\textwidth]{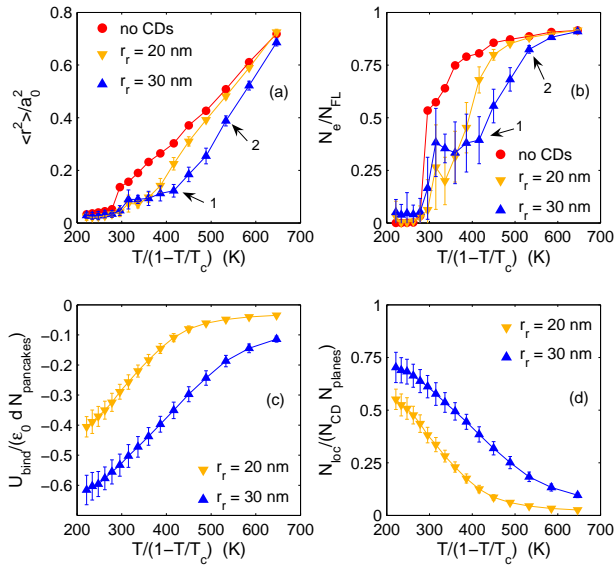}
\caption{(color online) Measured quantities
  for $B=100$ G. The pristine case and the case of CDs with radii of
  20 and 30 nm are shown. 
(a) Mean square transverse fluctuations of the FLs. For CDs of radius
30 nm the melting and delocalization transitions are indicated with
arrows. The first stage melting occurs for $C_L^2\approx 0.1$ where
$C_L$ is the Lindemann parameter.
(b) Number of loops of size greater than unity. A measure of FL entanglement.
(c) Binding energy per pancake between FLs and CDs.
(d) Number of capture pancakes per CD per plane.}
\label{figure2}
\end{figure}

In Fig.~\ref{figure2} we display different measurable quantities that are
indicators of the melting transitions. The red solid circles represent the
pristine system with no CDs and the melting appears sharp both for the
magnitude of the transverse fluctuations (a) and for the measure of
entanglement (b), given as the fraction of FLs not ending on
themselves by the periodic boundary conditions in the z-direction but
rather, due to the proliferation of permutations resulting in composite
loops made up of several individual segments. For the 30 nm CDs the
two stage melting is quite evident. For the transverse fluctuations we
first see a broad flat part ending in the first melting and then a
second kink signaling the approach of the fluctuations toward the pristine
value that indicates the melting of the skeleton. These transitions are
marked by the arrows. The flat part and the final kink are also
evident in Fig.~\ref{figure2} (b). For 20 nm the flat part is shorter and both
transitions occur closer to the pristine melting. In subplots (c) and
(d) we display the binding energy per pancake and the fraction of
trapped pancakes per CD per plane. Both these quantities become very
small at the delocalization transition. From here onward we limit
the discussions to CDs of radius of 30 nm.
\begin{figure}
\includegraphics[width=0.4\textwidth]{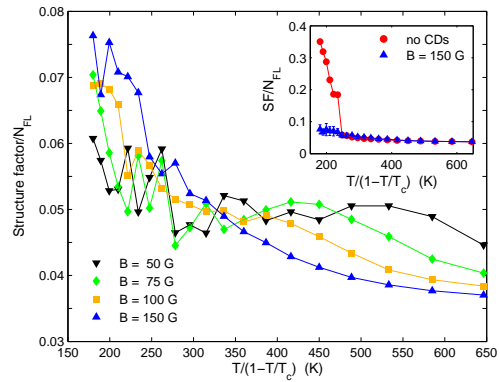}
\caption{(color online) Normalized structure factor at the first Bragg peak
  for different values of the magnetic field. 
The enhancement of the structure factor just below the field-dependent
melting transition is apparent. The inset
  shows the pristine vs. irradiated sample for $B=150$ G. }
\label{figure3}
\end{figure}
\begin{figure}
\includegraphics[width=0.4\textwidth]{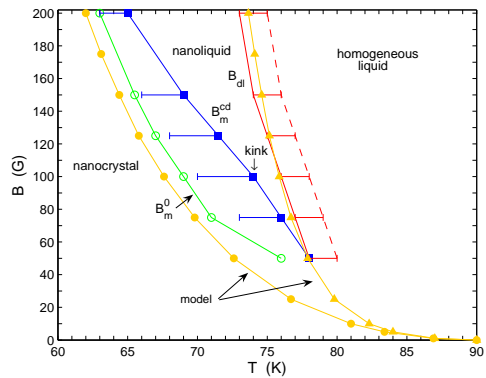}
\caption{(color online) Phase diagram in the $T-B$ plane as suggested
  from the results of the simulations.}
\label{figure4}
\end{figure}

In Fig.~\ref{figure3} we display the structure factor as a function of
the reduced temperature for  
different fields. The structure factors is not a good indicator of the
melting transition when CDs are present since the random skeleton make
it very low to begin with and the transition is broadened and becomes
second order or weakly first order. We can see clearly the rise of the
structure factor during the onset of the melting of the
nanocrystalline solid since it becomes easier for the interstitial
vortices to find better positions that are closer to hexagonal order
before the full melting occurs. We found evidence for this by looking
at snapshots at this temperature range. See e.g. Fig.~\ref{figure1}
(b), although there 
are more striking examples. A similar effect was discussed in
Ref.~\onlinecite{Tyagi} for weaker columnar pins of smaller radius. 
The inset shows the sharp pristine melting transition for $B=150$ G as
indicated by the step in the structure factor in contrast to the much weaker
decline in the case of columnar pins.

In Fig.~\ref{figure4} we display the phase diagram as it emerges from
the results 
of the simulations. The green line connecting the empty circles is the
location of the melting transition for the pristine system. The blue
line connecting the filled squares denote the melting of the
``irradiated'' systems when CDs of radius 30 nm are present. This line
denotes the conclusion of the melting of the nanocrystalline solid
which now becomes a nanoliquid. The horizontal blue bars denote the
broadening of the first stage melting characterized by the flat
sections in Fig.~\ref{figure2} (a) and (b). The two red lines (solid
and dashed) 
connected by horizontal error bars denote the second stage melting of
the skeleton which can be determined to a precision of about 2 K.
Note that for $B_\phi =50$ G the two transitions merge
together. The kink in the first-stage melting is defined by the point of
largest shift of the melting temperature from its pristine value and
this occurs at $B\approx 100$ G. A change in slope of the melting line
is observed there. In our simulation at this point the two melting
transitions do not yet coincide. This is similar to the experimental results
reported in Ref. \onlinecite{Banerjee2} for $B_\phi=60$ G. We also show in
Fig. 4 the prediction of a simple theoretical model described as follows:

The cage model is often used to get a rough estimate of the melting
transition\cite{Crabtree}. For a FL trapped by a CD we add the binding
energy between the FL and the 
CD to the harmonic cage potential. Thus the free-energy functional for
a single FL is given by:
\begin{eqnarray}
 F=\int_0^L dz
 \left[\frac{\epsilon_1}{2}\left(\frac{d\bm{r}(z)}{dz}\right)^2+U(\bm{r})+ 
\frac{1}{2}k \bm{r}^2\right],  \label{eq:model}
\end{eqnarray}
where $U(\bm{r})$ is the binding energy to the CD ($U(r)=V(r)/d$ as
given in Eq.(1) above), $\epsilon_1$ is the tilt modulus and $k\approx
\epsilon_0/a_0^2$ where $a_0=\sqrt{\phi_0/B}$ is the mean separation
between FLs. We now utilize the  
mapping of this model to a quantum particle in a potential where the
correspondence is $\hbar\rightarrow T$,
$m\rightarrow\epsilon_1$ and $z$ becomes the imaginary time.
In the limit $L\rightarrow \infty$ the particle is at zero temperature
and its energy is given by the ground state of the corresponding
Schrodinger equation
\begin{eqnarray}
 \left[-\frac{T^2}{2\epsilon_1}\bm{\nabla}^2+U(\bm{r})
+\frac{1}{2}k\bm{r}^2\right] \psi(\bm{r}) = E \psi(\bm{r}), \label{eq:shrod}
\end{eqnarray}
in two spatial dimensions. We have chosen the following value for
$\epsilon_1$
\begin{eqnarray}
  \label{eq:epsilon1}
  \epsilon_1\approx \epsilon_0 \left(\frac{1.55+\ln(\lambda/d)}
{\gamma^2}+\frac{d^2}{\lambda^2}\right).
\end{eqnarray}
The first term approximately represents the contribution of the
Josephson interaction 
\cite{Koshelev,Goldschmidt} and the second term results from the
electromagnetic 
interaction \cite{Clem}. We solved the Schrodinger equation numerically
using either the shooting method or alternatively the matrix
method. The equation reduces to a one-dimensional radial equation. The
melting transition is obtained 
by calculating the expectation value of $r^2$ and using the Lindemann
criterion with $C_L^2\approx 0.1$ to identify the melting
temperature. For $U=0$ we obtained the approximation to the pristine
melting curve displayed in Fig.~\ref{figure4}. Including $U(r)$ we obtain the
result for $B_{dl}$, the field corresponding to the skeleton
melting. We have also solved numerically the Schrodinger equation
without the cage and used the resulting energy and localization length
to reconstruct $B_{dl}$ following Lopatin and Vinokur \cite{Lopatin}. The result is
reasonable but the agreement is not as good than the method described
above. The delocalization field obtained from the simulation and from
the solution of the Schrodinger equation satisfies approximately
$B_{dl}\sim \exp(-T/T_0)$ with $T_0\approx 3$ K. This line is somewhat
steeper than the experimental result \cite{Banerjee2}, but this is also
true for the melting line that is flatter in the experiment. This is most
likely due to the fact that experimentally pristine samples contain a
certain amount of point defects that lower the melting temperature and
which are even more effective in lower temperatures and higher fields.

{\it Conclusions} - In this paper we used the multilevel Monte Carlo method to
simulate the pancake vortices in the highly anisotropic
superconductor BSCCO in the presence of columnar
defects of low concentration. When the applied field is larger than
the matching field we have found evidence for a sequence of two melting
transitions, one associated with the melting of crystallites and the
second with the demise of the skeleton - the FLs attached to the
CDs. Of course this is not a conclusive evidence that the latter
transition is a true thermodynamic transition. The agreement with
experiment is amazingly good for a system of 1296 pancakes when
averaging over 10 realizations of the disorder and display most of the
salient features found in the experiments. In addition we found a new
feature - a slight but noticeable enhancement of the structure factor
just prior to the melting transition when CDs are present. This
enhancement might 
be observed in experiments using small-angle neutron scattering
technique (SANS) which can effectively be used to measure the
structure factor.

{\it Acknowledgments} - This work is supported by the US Department of
Energy (DOE), Grant No. DE-FG02-98ER45686. Some of the simulations
were done at the Pittsburgh Supercomputing center under Grant
No. DMR950009P. YYG would like to thank Eli Zeldov for useful discussions
and for his hospitality at his lab at the Weizmann institute.
   
\end{document}